# GEOMETRY OF PSYCHOLOGICAL TIME[*]

METOD SANIGA

International Solvay Institutes for Physics and Chemistry,
Free University of Brussels (ULB), Campus Plaine, CP–231, Blvd du Triomphe,
1050 Brussels, Belgium
&
Astronomical Institute of the Slovak Academy of Sciences,
05960 Tatranská Lomnica, Slovak Republic

**Abstract:** The paper reviews the most illustrative cases of the "peculiar/anomalous" experiences of time (and, to a lesser extent, also space) and discusses a simple algebraic geometrical model accounting for the most pronounced of them.

## 1. Introduction

One of the most striking and persistent symptoms of so-called "altered" states of consciousness is, as we shall soon demonstrate, *distortions* in the perceptions of time and space. *Time* is frequently reported as flowing faster or slower, expanded or contracted, and may even be experienced as being severely discontinuous ("fragmented"). In extreme cases, it can stop completely or expand unlimitedly. The sense of space is likewise powerfully affected. *Space* can appear amplified or compressed, condensed or rarefied, or even changing its dimensionality; it can, for example, become just two-dimensional ("flat"), acquire another dimensions, or be reduced to a dimensionless point in consciousness.

As yet, there exists *no* mathematically rigorous and conceptually sound framework that would provide us with *satisfactory* explanations of these phenomena. Physics itself, although being the most sophisticated scientific discipline in describing the "objective" world, is not even able to account for the ordinary perception of time, let alone its other, more pronounced "peculiarities" mentioned above. Nor does it offer a plausible and convincing interpretation of the observed macroscopic dimensionality of space – giving more conceptual challenges than satisfactory answers. It was, among other things, this failure of current paradigms to accommodate a vast reservoir of the phenomena described above that originally motivated our search for a rigorous and self-consistent scheme, and which ultimately led into what we call the theory of *pencil-generated* space-times [1–9].

The aim of this contribution is to demonstrate that this theory represents a cogent starting point for a deeper understanding of the altered states of consciousness in their temporo-spatial aspects. It is shown, in particular, that the three most abundant groups of "pathological" perceptions of time, namely the feeling of *timelessness* ("eternity"), time *standing still*, as well as the experience of the *dominating past*, can well be modelled by singular (space-)time configurations represented by a specific pencil of conics. Being speculative, the paper is also offered to stimulate further research into the possible links between mathematics and physics on the one side, and psychology, psychiatry and philosophy on the other.

**2. Examples of Psychopathology of Time**

2.1. Near-Death Experiences

A typical near-death experience (NDE) occurs if a person is exposed suddenly to the threat of death but then survives and reports such phenomena as floating out of his/her body, moving rapidly through dark, empty space, having *the life review*, and encountering a brilliant white light. Out of these four consecutive phases it is the third one, the life review, which is of concern here. The following extract is taken from a famous book by R. Moody [10]:[1]

"After all this banging and going through this long, dark place, all of my childhood thoughts, *my whole entire life was* there at the end of this tunnel, just *flashing in front of me*. It was not exactly in terms of pictures, more in the form of thoughts, I guess. It was just *all there at once*, I mean, not one thing at a time, blinking on and off, but it was everything, *everything at one time*..."

However, it is not only the *past* but – weird as it may sound – also the *future* that a subject experiencing an NDE can have access to. The first to draw attention to this fact seems to have been K. Ring [11]:

"...the material I have collected that bears upon a remarkable and previously scarcely noted precognitive feature of the NDE I have called the personal *flashforward* (PF). If these experiences are what they purport to be, they not only have *extremely profound implications for our understanding of the nature of time* but also possibly for the future of our planet...
   Personal flashforwards usually occur within the context of an assessment of one's life during an NDE (i.e. during a life review and preview), although occasionally the PF is experienced as a subsequent vision. When it takes place while the individual is undergoing an NDE, it is typically described as an image vision of the future. *It is as though the individual sees something of the whole trajectory of his life, not just past events...* The understanding I have of these PFs is that to the NDEr they represent events of a *conditional future* – i.e., if he chooses to return to life, then these events *will* ensue..."

A more impressive description of this fascinating phenomenon is borrowed from [12], based on Atwater's personal experience:

"This time, I moved, not my environment, and I moved rapidly… My speed accelerated until I noticed a wide but thin-edged expanse of bright light ahead, like a "parting" in space or a "lip," with a brightness so brilliant it was beyond light yet I could look upon it without pain or discomfort… The closer I came the larger the parting in space appeared until… I was absorbed by it as if engulfed by a force field…
   Further movement on my part ceased because of the shock of what happened next. Before me there loomed two gigantic, impossibly huge masses spinning at great speed, looking for all the world like cyclones. One was inverted over the other, forming an hourglass shape, but where the spouts should have touched there was instead incredible rays of power shooting out in all directions… I stared at the spectacle before me in disbelief…
   As I stared, I came to recognize my former Phyllis self in the midupperleft of the top cyclone. Even though only a speck, *I could see my Phyllis clearly, and superimposed over her were all her past lives and all her future lives happening at the same time in the same place as her present life. Everything was happening at once! Around Phyllis was everyone else she had known and around them many others...* The

---

[1] In this and all the subsequent excerpts/quotations, italics are used to emphasize those parts of the narratives that most directly relate to the topic of the section. They are introduced by the author of the present paper, not the author(s) of the paper/book quoted.

*same phenomenon was happening to each and all. Past, present, and future were not separated but, instead, interpenetrated like a multiple hologram combined with its own reflection.*

The only physical movement anyone or anything made was to contract and expand. There was no up or down, right or left, forward or backward. There was only in and out, like breathing, like the universe and all creation were breathing – inhale/exhale, contraction/expansion, in/out, off/on."

2.2. Drug-Induced States

One of the most pronounced "distortions" in perception of time and space is encountered in the extraordinary states induced by the use of drugs. The following extract, taken from [13], illustrates this in detail:

"...This and all other changes in my dreams were accompanied by deep-seated anxiety and funeral melancholy, such as *are wholly incommunicable by words*. I seemed every night to descend – not metaphorically, but literally to descend – into chasms and sunless abysses, depths below depths, from which it seemed hopeless that I could ever re-ascend. Nor did I, by waking, feel that I had re-ascended. Why should I dwell upon this? For indeed the state of gloom which attended these gorgeous spectacles... *cannot be approached by words*.

The *sense of space*, *and* in the end *the sense of time*, *were both powerfully affected*. Buildings, landscapes, etc., were exhibited in proportions so vast as the bodily eye is not fitted to receive. *Space* swelled, and was *amplified to an extent of unutterable and self-repeating infinity*. This disturbed me much less than *the vast expansion of time*. Sometimes I seemed to have lived for seventy or a hundred years in one night; nay, sometimes had feelings representative of a *duration far beyond the limits of any human experience*..."

Here, one should notice that the "amplification" of space is often reported hand in hand with the "expansion" of time. Even a more dramatic and profound departure from the "consensus reality," induced by LSD, is depicted in [14]:

"... I found myself in a rather unusual state of mind; I felt a mixture of serenity and bliss... It was a world where miracles were possible, acceptable and understandable. I was *preoccupied with the problems of time and space and* the insoluble *paradoxes of infinity and eternity* that baffle our reason in the usual state of consciousness. I could not understand how I could have let myself be "brain-washed" into accepting the simple-minded concept of one-dimensional time and three-dimensional space as being mandatory and existing in objective reality. It appeared to me rather obvious that there are no limits in the realm of spirit and *that time and space are arbitrary constructs of the mind.* Any number of *spaces with different orders of infinities* could be deliberatery created and experienced. *A single second and eternity seemed* to be *freely interchangeable*. I thought about higher mathematics and saw *deep parallels between* various *mathematical concepts and altered states of consciousness*..."

This description clearly indicates that the mind is not confined to the limits of conventional space and time, and what we perceive in our "normal" state of consciousness is only a tiny fraction of the world we all have potential access to. The following experience of "disordered," "chaotic" time [15], induced by the drug called mescaline, dovetails nicely with the above statement:

"For half an hour nothing happened. Then I began feeling sick; and various nerves and muscles started twitching unpleasantly. Then, as this wore off, my body became more or less anaesthetized, and I became 'de-personalized', i.e., I felt completely detached from my body and the world…

This experience alone would have fully justified the entire experiment for me…, but at about 1.30 all interest in these visual phenomena was abruptly swept aside when I found that *time was behaving even more strangely than color*. Though perfectly rational and wide-awake… *I was not experiencing events in the normal sequence of time. I was experiencing the events of 3.30 before the events of 3.0; the events of 2.0 after the events of 2.45, and so on. Several events I experienced with an equal degree of reality more than once*. I am not suggesting, of course, that the events of 3.30 *happened* before the events of 3.0, or that any event *happened* more than once. All I am saying is that *I experienced them, not in the familiar sequence of clock time, but in a different, apparently capricious sequence which was outside my control.*

By 'I' in this context I mean, of course, my disembodied self, and by 'experienced' I mean learned by a special kind of awareness which seemed to comprehend yet be different from seeing, hearing, etc.… I count this experience, which occurred when, as I say, I was wide awake and intelligent, sitting in my own armchair at home, as *the most astounding and thought-provoking* of my life…"

The final experience we introduce in this section, induced and sustained by smoking *salvia divinorum*, seems to feature elements of time travel, or existing in separate realities simultaneously [16]:

"… the salvia started to overtake me. Suddenly, I was unsure of where I was and, more specifically, when I was. I wasn't sure if I was sitting on the floor in my new apartment or on the couch of my old one the previous week. *It felt as if I were in both places at once*, smoking salvia. *I felt I became unstuck in time. It seemed I was existing simultaneously in the past week's trip, the current moment, and thousands of other times, both in the future and the past. Not only other times of my life, but of other's lives as well, all existing as a four dimensional hyperbeing linked through salvia.* My vision had a very "edged" aspect, as if everything had an extra dimension. While I was laying on the floor with eyes closed, "time tripping", I didn't exactly see anything, but *I had a definite sense of being in numerous places, a sort of mental map*…

In all of my experiences, I get the impression that I am "bringing back" only a small portion of what I am experiencing. The sensations come at a breakneck pace, and it is difficult to even hang on, much less pay attention to what is actually going on. All of my experiences seem to have a somewhat consistent aspect. They feel *very real*, in a strange way…"

Note a striking resemblance between this experience and Atwater's NDE described in the previous section.

2.3. Mental Psychoses

This is the domain where much is still unsettled and uncertain and which thus provides an invaluable source for scientific imagination, as we strive to decipher the laws of Nature. In the accounts sampled below we shall recognise at least four distinct types of anomalous temporal patterns reported by mentally ill patients.

The first type is what the majority of psychotics refer to as "*time standing still.*" Some spectacular examples are found in a paper by H. Tellenbach [17], namely [17; p. 13]:

"I sure do notice the passing of time but couldn't experience it. I know that tomorrow will be another day again but don't feel it approaching. I can estimate the past in terms of years but I don't have any connection to it anymore. The time *standstill* is infinite, I live in a constant *eternity*. I see the clocks turn but *for me time does not flow*... Everything lies in one line, there are *no differences of depth* anymore... Everything is like a firm *plane*…"

and [*ibid*; p. 14]:

"Everything is very different in my case, time is passing very slowly. Nights last so long, one hour is as long as usually a whole day..." Sometimes *time* had totally *stood still*, it would have been horrifying. Even space had changed: "Everything is so empty and dark, everything is so far away from me... *I don't see space as usual*, I see everything as if it were just a background. It all seems to me *like a wall*, everything is *flat*. Everything presses down, everything looks away from me and laughs..."

It is worth noticing that when time comes to a stillstand, perceived space loses one dimension, becoming only two-dimensional. We shall see later that this feature finds a very nice explanation in our model. A slightly more detailed description of this temporal mode is given in a very readable paper by Muscatello and Giovanardi Rossi [18; p. 784]:

"*Time is standing still for me*, I believe. It is perhaps only a few moments that I have been so bad. I look at a clock and I have the impression, if I look at it again, that an *enormous period of time* has passed, as if hours would have passed instead only a few minutes. It seems to me that a duration of time is enormous. *Time does not pass any longer, I look at the clock but its hands are always at the same position, they no longer move, they no longer go on; then I check if the clock came to a halt, I see that it works, but the hands are standing still*. I do not think about my past, I remember it but I do not think about it too much. When I am so bad, *I never think about my past*. Nothing enters my mind, nothing... I did not manage to think about anything. I did *not* manage to see *anything in my future. The present does not exist for me* when I am so bad... *the past does not exist, the future does not exist*."

The second type of temporal psychosis is what one may well call the *dominating past*. A couple of examples below, both by schizophrenics, give detailed accounts of it. The first narrative makes explicit how the temporal is devoid of the notions of both the future and the present [19; p. 563]:

"I stop still, I am being thrown *back into the past* by words that are being said in the hall. But this all is self evident, it must be that way! There is *no present* anymore, there is *only* this stated *being related to the past*, which is *more than a feeling*, it goes through and through. There are all sorts of plans against me in the air of this hall. But I don't listen to them, I let my mind rest so that it doesn't corrode... Is there any future at all? Before, the future existed for me but now it is shrinking more and more. *The past is so very obtrusive, it throws itself over me; it pulls me back...* By this I want to say that there is *no future* and I am thrown back... Strange thoughts enter my mind and drive me off into the past..."

The other account seems to even question the very nature of time [*ibid*.; p. 561]:

"It pulls me back, well, where to? To where it comes from, there, where it was before. *It enters the past*. It is that kind of a feeling as if you had to fall back. This is the disappearing, the vanishing of things. *Time slips into the past*, the walls are fallen apart. Everything was so solid before. It is as if it were so close to be grabbed, as if you had to pull it back again: *Is that time?* Shifted way back!"

The third characteristic type of distorted temporal dimension a psychotic often encounters is the sensation of time flowing *backward*. Of all the psycho-time-related references we have seen, no account draws a portrait of the essential properties of this mode better than that found in [19; p. 556]:

"Yesterday at noon, when the meal was being served, I looked at the clock: why did no one else? But there was something strange about it. For the clock did not help me any more and did not have anything to say to me any more. How was I going to relate to the clock? *I felt as if I had been put back, as if something of the past returned, so to speak, toward me, as if I were going on a journey. It was as if at 11:30 a.m. it was 11:00 a.m. again, but not only time repeated itself again, but all that had happened for me during that time as well*. In fact, all of this is much too profound for me to express. In the middle of all this something happened which did not seem to belong here. *Suddenly, it was not only 11:00 a.m. again, but a time which passed a long time before was there and there inside* – have I already told you about a nut in a great, hard shell? It was like that again: *in the middle of time I was coming from the past towards myself*. It was dreadful. I told myself that perhaps the clock had been set back, the orderlies wanted to play a stupid trick with the clock. I tried to envisage time as usual, but I could not do it; and then came a feeling of horrible expectation that *I could be sucked up into the past, or that the past would overcome me and flow over me*. It was disquieting that someone could play with time like that, somewhat daemonic..."

A psychotic patient of Laing [20] gives a very brief and concise description:

"... *I got the impression that time was flowing backward; I felt that time proceeded in the opposite direction*, I had just this extraordinary sensation, indeed... the most important sensation at that moment was, *time in the opposite direction*... The perception was so real that I looked at a clock and, I do not know how, I had the impression that the clock confirmed this feeling, although I was not able to discern the motion of its hands..."

A strikingly similar portrayal of time-reversal is also provided by a depressive patient of Kloos [21; p. 237]:

"As I suddenly broke down I had this feeling inside me that time had completely flown away. After those three weeks in a sick-camp, I had this feeling that the clock hands run idle, that they do not have any hold. This was my sudden feeling. I did not find, so to speak, any hold of a clock and of life anymore, I experienced a dreadful psychological breakdown. I do not know the reason why I especially became conscious of the clock. *At the same time, I had this feeling that the clock hands run backward...* There is only one piece left, so to speak, and that stands still. I could not believe that time really did advance, and that is why I thought that the clock hands did not have any hold and ran idle... *As I worked and worked again, and worried and did not manage anything, I simply had this feeling that everything around us (including us) goes back... In my sickness I simply did not come along and then I had this delusion inside me that time runs backward...* I did not know what was what anymore, and I always thought that I was losing my mind. *I always thought that the clock hands run the wrong way round, that they are without any meaning. I just stood-up in the sick-camp and looked at the clock – and it came to me then at once: well, what is this, time runs the wrong way round?!... I saw, of course, that the hands moved forward, but, as I could not believe it, I kept thinking that in reality the clock runs backward...*"

The final type of temporal psychosis can be termed the *extended present*, and is described nicely in [22; pp. 104–107]:

"The *patient elevates* herself *above normal boundaries of time* without totally surmounting them. The distinction of the present and the future is not cancelled out as the patient still speaks about both dimensions, yet *the line between* the actual *present and* the only maybe-possible and unreal *future becomes swaying and possible to cross*. Both dimensions incapsulate and overlap each other without a steady transition. *The future fuses with the present* and vice versa and experiencing acquires a flickering twilight character which is radically distinguished from how a healthy person anticipates the future in day-dreams and the like... *The edge between the present and the past is swaying as well*. At the same time and in a totally different way, *the past is included in and fuses with the events of the present* as well as usually the present is part of the past. There is a kind of *condensation of time*; the present is not distinguished amidst the continuous, steady flow of the past any more, but at the same time the present is not filled with something past as it usually is with normal people; in this case it overlaps...

*The three temporal levels* of past, present and future therefore *seemed to overlap* in the psychotic experience of the patient *in an extremely peculiar simultaneousness* without really invalidating the distinction of past, present and future."

## 2.4. Mystical States

In the last example the present loses its "point-like" nature and starts to expand into *both* the future *and* the past. If this expansion is not limited, the experiencer will eventually attain the state of *the all-containing present* ("eternity"), when he/she is able to see all events simultaneously, as in the following remarkably vivid narrative [23]:

"…I get up and walk to the kitchen, thinking about what a timeless experience would be like. I direct my attention to everything that is happening at the present moment – what is happening here, locally, inside of me and near me, but non-locally as well, at ever increasing distances from me. I am imagining everything that is going on in a slice of the present – throughout the country, the planet, the universe. It's all happening at once. *I begin to collapse time, expanding the slice of the present, filling it with what has occurred in the immediate "past."* I call my attention to what I just did and experienced, what led up to this moment, locally, but keep these events within a slowly expanding present moment. *The present slice of time slowly enlarges, encompassing, still holding, what has gone just before, locally, but increasingly non-locally as well.* By now, I am standing near the kitchen sink. The present moment continues to grow, expand. *Now it expands into the "future" as well*. Events are gradually piling up in this increasingly larger moment. What began as a thin, moving slice of time, is becoming thicker and thicker, increasingly filled with events from the "present,""past," and "future." *The moving window of the present becomes wider and wider, and moves increasingly outwardly in two temporal directions at once.* It is as though things are piling up in an ever-widening present. The "now" is becoming very thick and crowded! *"Past" events do not fall away and cease to be; rather, they continue and occupy this ever-widening present. "Future" events already are, and they, too, are filling this increasingly thick and full present moment. The moment continues to grow, expand, fill, until it contains all things, all events. It is so full, so crowded, so thick,* *that everything begins to blend together. Distinctions blur. Boundaries melt away. Everything becomes increasingly homogeneous, like an infinite expanse of gelatine. My own boundaries dissolve. My individuality melts away. The moment is so full that there no longer are separate things. There is no-thing here. There are no distinctions.* A very strong emotion overtakes me. Tears of wonder-joy fill my eyes. This is a profoundly moving experience. Somehow, I have moved away from the sink and am now several feet away, facing in the opposite direction, standing near the dining room table. *I am out of time and in an eternal present. In this present is everything and no-thing. I, myself, am no longer here. Images fade away. Words and thoughts fade away. Awareness remains, but it is a different sort of awareness. Since distinctions have vanished, there is nothing to know and no one to do the knowing. "I" am no longer localized, but no longer "conscious" in the usual sense. There is no-thing to be witnessed, and yet there is still a witnesser.* The experience begins to fade. I am "myself" again. I am profoundly moved. I feel awe and great gratitude for this experience with which I have been blessed…"

A somewhat bizarre, yet more scholarly report of an almost identical psycho-space-time pattern is found in [24]. Although the author almost exclusively focuses on the spatial fabric of existence throughout the book, it is undoubtedly the temporal aspect of this particular experience that is most fascinating:

"I woke up in a whole different world in which the puzzle of the world was solved extremely easily in a form of *a different space*. I was amazed at the wonder of this different space and this amazement concealed my judgement, *this space is totally distinct from the one we all know*. It had *different dimensions*, *everything contained everything else*. I was this space and this space was me. The outer space was part of this space, I was in the outer space and the outer space was in me...

Anyway, I didn't experience time, time of the outer space and aeons until the second phase of this dream. In the cosmic flow of time you saw worlds coming into existence, blooming like flowers, actually existing and then disappearing. It was an endless game. *If you looked back into the past, you saw aeons, if you looked forward into the future there were aeons stretching into the eternity*, and *this eternity was contained in the point of the present*. One was situated in a state of being in which the "will-be" and the "vanishing" were already included, and this "being" was my consciousness. It contained it all. This "being-contained" was presented very vividly in a geometric way in form of *circles of different size which* again were all part of a unity since all of the circles *formed exactly one circle. The biggest circle was part of the smallest one and vice versa*. As far as the differences of size are concerned, I could not give any accurate information later on..."

Note a striking similarity between this experience and the experience of Grof's subject (Sec. 2.2); in particular, both the subjects speak about the puzzling equivalence between the eternity and the moment of the present. This seems to be a very important property of a mystical state, for it is also mentioned by such famous mystics as St. Thomas and Meister Eckhart, and even by the great Dante Alighieri, as pointed out by Ananda Coomaraswamy [25; p. 110]:

"[*St. Thomas*:] Eternity is called "whole" not because it has parts, but because it is wanting in nothing... The expression "simultaneously whole" is used to remove the idea of time, and the word "perfect" to exclude the now of time... *The now that stands still is said to make eternity*..."

[*ibid*; p. 117]:

"[*Meister Eckhart*:] God is creating the whole world now, this instant ("nu alzemale")... He makes the world and all things in this *present Now* ("gegenwuertig nu")..."

[*ibid*; pp. 120–121]:

"Dante, when he is speaking of Eternity, makes many references to this "essential point" or "moment." *All times are present to it* ("il punto a cui tutti li tempi son presenti," Paradiso 17.17); *there every where and every when are focused* ("dove s'appunta ogni ubi ed ogni quando," Paradiso 29.12)... In it alone is every part there where it ever was, for it is not in space, nor hath it poles ("in quella sola è ogni parte là ove sempr'era, perché non è in loco, e non s'impola")... Whereby it thus doth steal from thy sight (22.64)."

## 3. An Outline of the Algebraic Geometrical Model of Time Dimension

In what follows we shall introduce the basic features of a simple geometrical model capable of mimicking remarkably well some of the most pronounced pathologies of time and space we highlighted in the preceding section. The presentation will be rather illustrative, so as to be accessible to scholars of various disciplines and diverse mathematical backgrounds. The reader wishing to go further into the details of the mathematical formalism is referred to our papers [3–5,8].

The model in question is based on a specific *pencil* (i.e. a linear, single parametric aggregate) *of conics* in the real projective plane and its structure is illustrated in Fig. 1. We see that all the conics touch each other in two different points, $B_1$ and $B_2$, and the corresponding two common tangent lines meet at the point S. This pencil of conics is taken to generate the *time* dimension, where each conic represents a single event. The pencil, as it stands, is homogeneous in the sense that every conic has the same footing in it. Yet, from what we have just seen it is obvious that the intrinsic structure of subjective time is far from being homogeneous, being, in fact, endowed with three different kinds of event, namely the past, present and future. Hence, the pencil has to be "de-homogenized" in order to yield the structure required.

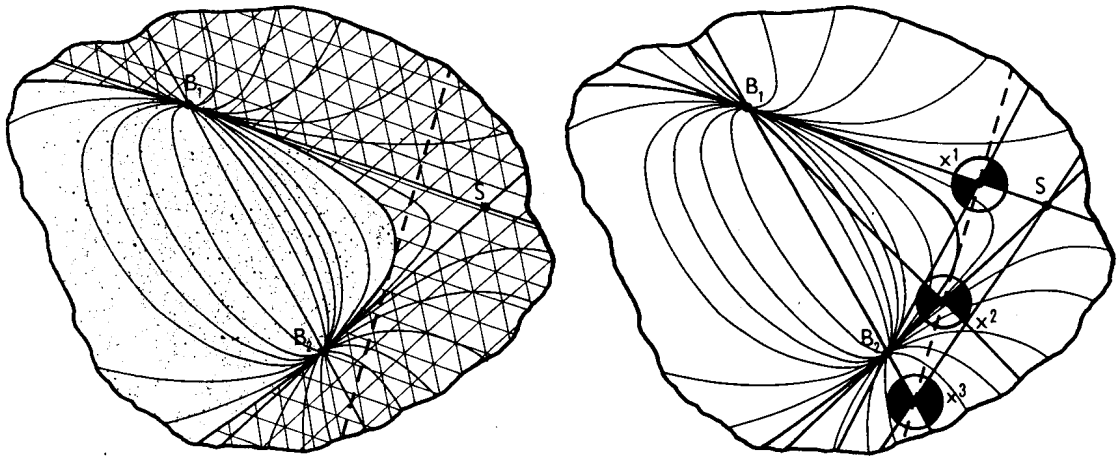

**Figure 1**.

This can be done fairly easily if, for example, we select in the plane one line (the broken line in Fig. 1) and attach to it a special status. It is clear that if this distinguished line is in a general position, it does not pass via any of the points $B_1$, $B_2$, and S. Under such an assumption, the conics of the pencil are seen to form, as far as the intersection properties are concerned, two distinct domains with respect to this line (see Fig. 1, *left*). One domain comprises those conics that have no intersection with the line (these conics are located in the dotted area and we shall call them "non-cutters"), whereas the other domain features the conics cut by the line in two distinct points (they are located in the shaded area and called "cutters"). These two domains are separated from each other by a *unique* single conic (drawn bold in Fig. 1) which has the broken line for its tangent (this conic will be referred to as the "toucher"). This is really a very remarkable pattern for it is seen to reproduce strikingly well, at least at the qualitative level, our ordinary perception of time once we postulate that the cutters represent the past events, the non-cutters the moments of the future, and that the unique toucher stands for the present, the now [1–9].

As for a *spatial* dimension, this will be modelled by a *pencil of lines,* i.e. by all the lines that pass through a *given* point (called the vertex of the pencil). Here the given point means any point which our broken line shares with each of the lines $B_1B_2$, $B_1S$, and $B_2S$, defined by the pencil of conics. From Fig. 1, *right*, it is evident that for a general position of the broken line there are just *three* specific pencils of lines (depicted in Fig. 1, *right*, as three half-filled circles). And there are just *three* spatial dimensions ($x^1$, $x^2$ and $x^3$) we perceive in our "normal/ordinary" state of consciousness! The model is thus characterized by an intricate connection between the *intrinsic* structure of time and the *number/multiplicity* of macroscopic spatial dimensions [2,3,8].

In order to make this link visible to the eye, let us start moving the broken line from its original, generic position of Fig. 1 towards the point S in such a way that it is eventually incident with the latter – as shown in Fig. 2. As it can easily be discerned from this figure, in this limiting case the toucher disappears and we find only the cutters (shaded area) and non-cutters (dotted area). In other words, our time dimension now *lacks* the moment of *the present*, being endowed with the past and future events only. As it is intuitively obvious that out of the three temporal levels, i.e. the past, present and future, it is the present that seems to be fully "responsible" for what we experience as the "flow/passage" of time, its absence in the above-mentioned arrow implies that such time does *not* pass, it *stands still*. From Fig. 2 it can further be discerned that this partial "collapse" of the generic arrow of time is accompanied by a *3⇒2* reduction in the dimensionality of space, because two of its coordinates ($x^1$ and $x^2$) merge with each other and form a single coordinate. This configuration thus bears a striking similarity to the space-time construct that a couple of Tellenbach's melancholic patients were trying to describe (see the first two excerpts in Sec. 2.3)!

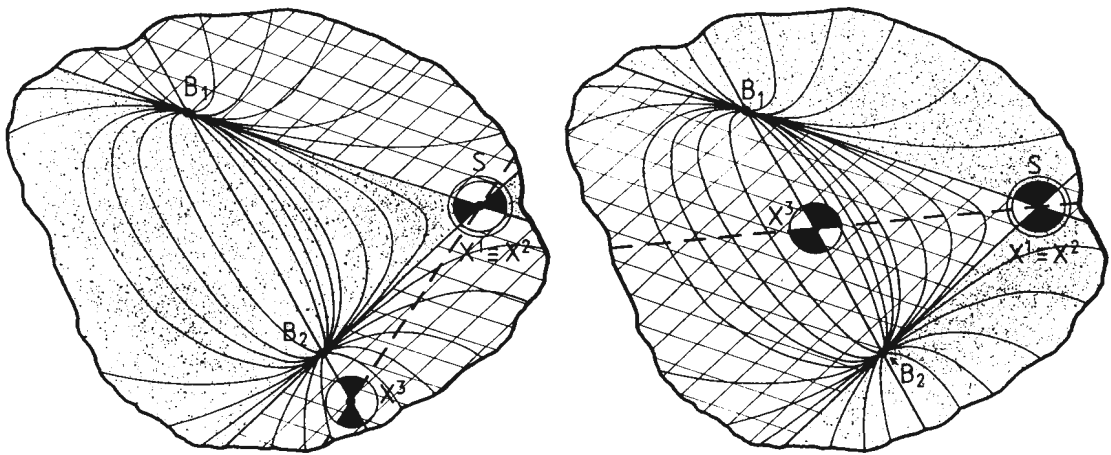

**Figure 2.**

Another kind of "degenerate" temporal arrow emerges when the broken line hits one of the points $B_1$, $B_2$, but does not incorporate the point S – the mode depicted in Fig. 3. It is obvious that the line selected in this way is a secant to *every* conic of the pencil, which means that the corresponding time dimension features *exclusively* the region of *the past* – that is, it is identical with the *dominating past* mode of F. Fischer's schizophrenic patients (see the fourth and fifth excerpts in Sec. 2.3). Note that there is again the $3 \Rightarrow 2$ drop in the number of space dimensions.

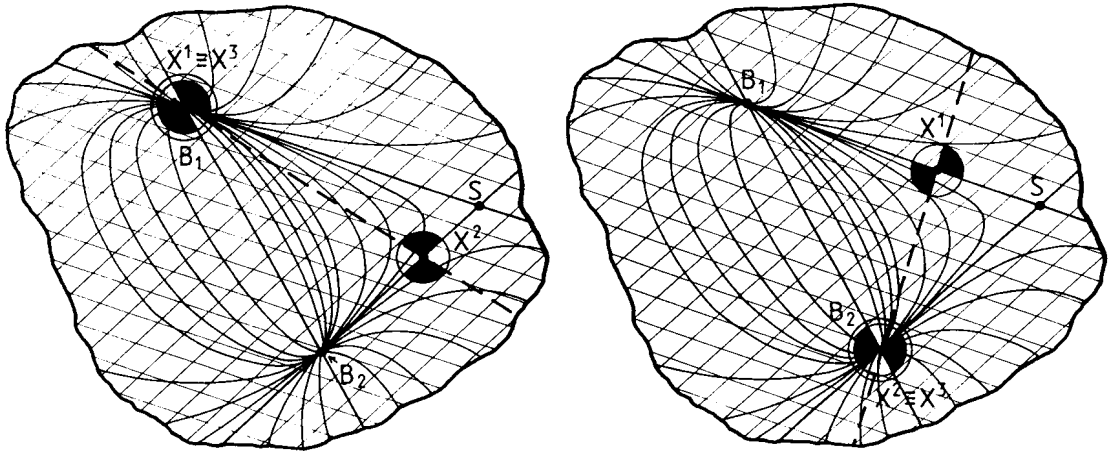

**Figure 3.**

The third, and the last fundamental mode associated with this particular pencil of conics, is characterized by the broken line coinciding with one of the common tangent lines, $B_1 S$ or $B_2 S$ – as shown in Fig. 4. In this case, *every* point of the broken line is the vertex of the pencil of lines representing a spatial coordinate, thus space becomes *infinitely* dimensional (which is illustrated in Fig. 4 by two lines running parallel to the line $B_1 S$ resp. $B_2 S$). On the other hand, the broken line is now tangent to *every* conic of the pencil, i.e. *all* the conics are its *touchers*; the corresponding time dimension thus consists *solely of* the events of *the present*, and represents thus nothing but what in the previous section was referred to as the "eternity." We see that this (kind of) space-time configuration possesses all the basic attributes of the space-time of mystics (see Sec. 2.4), as described by Huber's narrative.

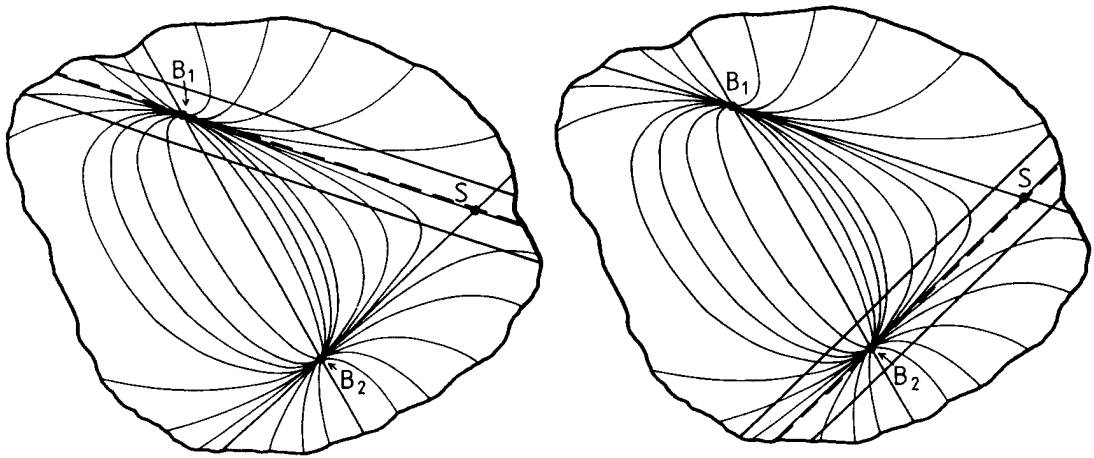

**Figure 4.**

The attentive reader may ask, "What kinds of (space-)time configurations do we find if we consider other types of a pencil of conics residing in the real projective plane?" As there exist as many as nine different types of these [3], we do find some new temporal patterns not exhibited by the previously discussed pencil. Thus, for example, we arrive at internally "contorted" forms of the time dimension, such as the one composed of two distinct domains of the past and two distinct moments of the present, but only *one* region of the future; or that endowed with two different domains of the past separated from each other by a single moment of the present [3]. In both the cases, the corresponding psychopathological counterparts have yet to be discovered.

A whole new class of temporal structures is revealed if we relax the assumption that the projective plane is *real* and consider also projective planes defined over other ground fields [4–6,8]. Thus, for example, we find that if the ground field is algebraically closed, the corresponding time dimension is always devoid of the concept of the future, irrespectively of both the type of pencil employed and the position of the distinguished (broken) line [4]. Even more intriguing is the case of so-called Galois (or finite) fields: here, the time dimension consists of finite numbers of events only, lacks any ordering (compare with the experience of "disordered" time of Sec. 2.2) and may even become transmuted into (indistinguishable from) a *spatial* dimension if these fields are of characteristic two [5,6].

Finally, a very promising generalization of the above-discussed rudimentary model is achieved if the constraint of the *linearity* of the aggregate of conics is also removed. A simple, quadratic set of conics put forward in [9] not only reproduces all the key features of the linear model, but it also leads to what we termed the *arrow-within-arrow* patterns – the structures accounting, for example, for experiences of time flowing *backward* (see Sec. 2.3). As we do not have additional space here to discuss these and other intriguing cases in further mathematical detail, the interested reader is referred to our papers [2–6,8,9].

## 4. Conclusion

The findings and results just described provide us with strong evidence that *not only* are the manifestations of mental or psychological time so diverse and unusual that they fail to conform to the generally adopted picture of the macroscopic physical world, *but* there also exists a unique mathematical framework which, at least qualitatively, underlies and unifies their seemingly bizarre properties. Hence, any attempt to disregard these psychopathological temporal constructs as pure hallucinatory phenomena would simultaneously cast a doubtful eye on the very role of mathematics in our understanding of Nature. To the contrary, it is just mathematics (algebraic projective geometry here) that plainly tells us that it is far more natural to expect all these "unusual" perceptions of time to be simply as real as our ordinary ("normal") one. We are, however, fully aware that this point of view is very likely to meet with scepticism, and even with fierce opposition, from the side of 'hard-line' instructional scientists. Most such scientists will probably object to the *anecdotal* character inherent in describing the variety of time's multifaceted phenomena. However, this inevitable anecdotal feature is necessary for research on the qualitative aspects of time – research which profits both psychology and physics. As very well argued by Shallis [26; p. 153]:

"*Quality* and *quantity* are somewhat like the ingredients of *descriptive* and *instructional* science, respectively. Because the two approaches are so different the sorts of evidence employed in each will also differ. *Whereas the instructional approach requires, indeed demands, rigorous, quantitative and reproducible evidence, the descriptive attitude, which often deals with the unique and individual, is mainly anecdotal*. This does not mean it is uncritical or sloppy, but in trying to find the whole truth everything must be taken into account. If some evidence turns out to be false, that too is part of the picture. In instructional science anecdotal evidence, *even if true*, can be dismissed as unquantifiable and impossible to assess. *The techniques of instructional science cannot handle individual experience or admit to the quality of time. Descriptive science can...*"

We are firmly convinced that anything that shows a definite mathematical structure, whatever bizarre and counter-intuitive it may appear, deserves effort and ingenuity to be thoroughly explored and examined, all the more that [*ibid*, pp. 153–4]:

"…the fact that the experience of time is *not* quantifiable puts it into arena of human perceptions that are at once richer and more meaningful than are those things that are merely quantifiable… *The lack of quantification of temporal experiences is not something that should stand them in low stead, to be dismissed as nothing more than fleeting perceptions or as merely anecdotal; rather that lack should be seen as their strength*. It is because the experience of time is not quantifiable and not subject to numerical comparison that makes it something of quality, something containing the essence of being…"

**Acknowledgements**

I am very grateful to Mr. Pavol Bendík for painstaking drawing of the figures. I would like to express my cordial thanks to Miss Daniela Veverková and Mr. Peter Hahman for translating into English all the excerpts taken from journals written in German. My warm thanks are due also to Dr. Rosolino Buccheri (IASFC, Palermo) for the corresponding translation of a couple of excerpts in Italian. I am also indebted to Prof. Mark Stuckey (Elizabethtown College) for a careful proofreading of the paper. Last, but not least, I wish to express my gratitude to my wife for her continuous support and encouragement of my work. This work was supported in part by the NATO Collaborative Linkage Grant PST.CLG.976850, the NATO Advanced Research Fellowship distributed and administered by the Fonds National de la Recherche Scientifique, Belgium, and the 2001–2003 Joint Research Project of the Italian Research Council and the Slovak Academy of Sciences "The Subjective Time and its Underlying Mathematical Structure."